\documentclass[10pt,conference]{IEEEtran}

\usepackage{cite}
\usepackage{amsmath,amssymb,amsfonts}
\usepackage{algorithmic}
\usepackage{graphicx}
\usepackage{textcomp}
\usepackage{xcolor}
\usepackage{tikz}
\usepackage{quantikz}
\usepackage{subcaption} 
\def\BibTeX{{\rm B\kern-.05em{\sc i\kern-.025em b}\kern-.08em
    T\kern-.1667em\lower.7ex\hbox{E}\kern-.125emX}}

\usepackage[notrig]{physics} 
\begin{document}

\title{Towards a Digital Twin of Noisy Quantum Computers: Calibration-Driven Emulation of Transmon Qubits}

\author{
Ronny Mueller\IEEEauthorrefmark{1}, Maximilian Zanner\IEEEauthorrefmark{1}, Mika Schielein\IEEEauthorrefmark{1}, Martin R\"{u}fenacht\IEEEauthorrefmark{1}, David Rabanus\IEEEauthorrefmark{3}, Eduardo Sch{\"a}tzle\IEEEauthorrefmark{3},\\ Kristel Michielsen\IEEEauthorrefmark{3}, Ashwin Kumar Karnad\IEEEauthorrefmark{3}, Dennis Willsch\IEEEauthorrefmark{3}, Elise Jennings\IEEEauthorrefmark{1}, Cica Gustiani\IEEEauthorrefmark{2} \\
\IEEEauthorrefmark{1}\textit{ParTec AG, Possartstraße 20, 81679 München, Germany}\\
\IEEEauthorrefmark{2}LiP6, \textit{Sorbonne Université}, Paris, France\\
\IEEEauthorrefmark{3} J{\"u}lich Supercomputing Centre, Forschungszentrum J{\"u}lich, J{\"u}lich, Germany\\
Corresponding authors: \texttt{Ronny.Mueller@par-tec.com, Cica.Gustiani@lip6.fr}
}

\maketitle

\begin{abstract}  
We develop a parametric error model to construct a digital twin of a superconducting transmon qubit device. The model parameters are extracted from hardware calibration data and supplementary benchmarking circuits, providing a dynamic, system-specific representation of noise and gate imperfections. Given the strong dependence of qubit performance on calibration procedures, our approach captures real-time device fluctuations. By incorporating predominant noise sources derived from underlying physical processes, we enhance the emulation's accuracy while reducing the data required for model fitting. Finally, we validate our model by comparing its predictions with experimental results from a 5-qubit QPU, achieving a mean total variation distance of 0.15 between the shot distributions. This digital twin can be leveraged for predictive performance analysis, error mitigation strategies, and the optimization of quantum protocols, contributing to more reliable quantum computations.
\end{abstract}

\begin{IEEEkeywords}
emulator, transmon qubits
\end{IEEEkeywords}

\section{Introduction}

Quantum computing has the potential to revolutionize problem-solving across multiple scientific domains, but current hardware is constrained by noise and errors that limit computational accuracy~\cite{Preskill_2018}. Noisy intermediate-scale quantum (NISQ) devices suffer from various imperfections, including gate errors, readout inaccuracies, decoherence, and crosstalk, all of which degrade algorithmic performance~\cite{Preskill_2018, zimborás2025mythsquantumcomputationfault}. While quantum error correction offers a long-term path toward fault-tolerant computing, its feasibility remains restricted by hardware limitations, despite recent advances~\cite{acharya2024quantumerrorcorrectionsurface}. In the near term, improving the understanding and mitigation of noise is essential for maximizing the performance of quantum circuits and enabling meaningful computations~\cite{Cai_2023}.

Noise in quantum hardware is not uniform; different sources contribute to circuit degradation in distinct ways. Crosstalk, for instance, disrupts the independence of quantum operations, leading to unintended interactions between qubits~\cite{10483460, Zhao_2022}. Similarly, frequency drift, control imperfections, and hardware-specific noise effects complicate the implementation of high-fidelity quantum gates~\cite{Georgopoulos_2021}. Accurate noise models must therefore capture these non-idealities at both the gate and circuit levels. While individual gate characterizations provide useful information for noise modeling, the behavior of relevant application circuits often still deviates from idealized assumptions, requiring more comprehensive modeling strategies~\cite{PhysRevResearch.5.043210}.

Superconducting qubits have become a leading platform for realizing state-of-the-art quantum processors. Standard cleanroom fabrication techniques allow for precise control over qubit parameters and a high degree of engineerability, enabling scalable integration~\cite{PhysRevLett.89.117901, Wendin_2017}. Another significant advantage of superconducting qubits lies in their capability for fast gate operations, typically in the nanosecond regime, enabling the execution of large quantum circuits~\cite{Arute2019}. Moreover, recent years have witnessed remarkable improvements in qubit coherence times, increasing from mere nanoseconds to millisecond timescales, driven by advances in materials engineering, device geometry optimization, and improved fabrication processes~\cite{Paik_2011, Anderson2020, bland20252dtransmonslifetimescoherence}. Despite these substantial achievements, managing noise -originating from thermal fluctuations, electromagnetic interference, and inter-qubit crosstalk- remains critical as quantum processors continue to scale~\cite{Krantz_2019, Preskill_2018}.

Experimental techniques such as quantum process tomography (QPT)~\cite{Gross:2013hri, Altepeter_2003} and gate set tomography (GST)~\cite{greenbaum2015introductionquantumgateset, Nielsen_2021} have been developed to characterize noise in quantum processors. Applying these methods at scale, however, remains challenging due to the computational cost of characterizing large quantum systems. Integrating these characterizations into practical noise models is crucial for improving the predictive accuracy of quantum circuit simulations and guiding the development of noise-resilient algorithms. Significant research effort has already been put towards this goal. This includes detailed works on single physical noise sources and their characterization, both, platform-specific~\cite{Dai_2021, Ketterer_2023, sete2024errorbudgetparametricresonance} and platform-independent~\cite{Gambetta_2012, Sarovar_2020}. Further noise model proposals rely to different degrees on fitting fully or partially parametrized error models~\cite{weber2023volumetricbenchmarkingquantumcomputing, PhysRevA.103.042603, Harper_2020}, where our selection is focusing on qubits that are based on superconducting technology.

However, the noise models presented above incorporate many errors that are not directly derived from physical principles but rely on resource-intensive benchmarking experiments and/or require the fitting of a large number of parameters. In contrast, we introduce a noise model based on the underlying physical characteristics of a fixed-coupling, flux-tunable superconducting transmon device. Our approach primarily leverages qubit properties that are routinely obtained from standard calibration procedures, reducing the dependence on extensive additional benchmarking. For a $n$-qubit device, our model requires fitting only $n$ parameters, making it both computationally efficient and scalable while maintaining interpretability in its predictions. The emulator is designed to closely model the behavior of superconducting qubits by incorporating parameters such as relaxation and dephasing times, gate fidelities, and readout errors, which are routinely extracted during qubit calibration. By being fed with experimentally relevant data, the emulator provides a reliable and practical tool for optimizing quantum control strategies and refining quantum hardware designs.

\section{Parameterized noise model} \label{sec:noise_model_theo}

\noindent In this section, we describe the noise model
using different error types, along with the underlying assumptions. We specify the noise parameters and detail the methods used to obtain them. Our model follows a similar structure to that in reference~\cite{Gustiani2025virtualquantum}. Before presenting the details, we outline the general assumptions underlying our model as follows.

First, any operator or noise acting on a qubit is subjected to the rotating-wave approximation and is implemented within its own rotating frame of reference. In the absence of gate application or noise, the qubit undergoes an identity operation. For two-qubit interactions or noise, we approximate their dynamics using an effective reference frame. Second, the evolution of the quantum state is discretized using first-order Trotterization, which defines the corresponding error bound. Third, we model the application of a noisy unitary operation $U$ as $\mathcal E\circ U$, where $\mathcal E$ is a completely positive trace-preserving (CPTP) map, as we consider only Markovian noise. Finally, the noise model considers an array of different noise sources, described in the following.

\paragraph{Reset error} Resetting the qubits is modeled as their relaxation to the thermal equilibrium state $\rho_T$ at a finite small temperature. In practice, this can be implemented using thermal relaxation or an active reset method~\cite{Magnard_2018}. 
Thus, noisy preparation of the state zero 
$\ket{00\dots0}$ is modeled as
\begin{equation}
\label{eq:rhot}
\rho_T=\bigotimes_{j}(1-p^1_j)\ketbra0 + p^1_j\ketbra1, 
\end{equation}
where $p^1_j$ indicates the probability of being in the excited state when initializing qubit $j$. 
The state $\rho_T$ is obtained by applying the (generalized) amplitude damping channel, denoted as $gamp(\gamma)$, with the following Kraus operator:
\begin{equation} 
\begin{aligned}
    \label{eq:gdamp}
gamp&(\gamma) =
\Biggl\{
    \sqrt{1-p^1_j}\begin{pmatrix}
        1 & 0 \\
        0 & \sqrt{1-\gamma}
    \end{pmatrix},
    \sqrt{1-p^1_j}\begin{pmatrix}
        0 & \sqrt{\gamma} \\
        0 & 0
\end{pmatrix}, %
\\ &
\sqrt{p^1_j}
    \begin{pmatrix}
        \sqrt{1-\gamma} & 0 \\
        0 & 1
    \end{pmatrix}
,
 \sqrt{p^1_j}\begin{pmatrix}
        0 & 0\\
        \sqrt{\gamma} & 0
\end{pmatrix}
\Biggl\},
\end{aligned}
\end{equation}
where the parameter $\gamma \in [0,1]$ controls the strength of the damping, and for $\gamma = 1$, the channel maps an arbitrary state to $\rho_T$. This model requires parameters of the fraction of the population in the excited state for each qubit $\{p^1_j\}$.

\paragraph{Exponential decay error}

The coherence time of qubits limits the capability to execute deep circuits. This notion of coherence is commonly characterized by the relaxation times $T_1$ and $T_2$, which represent the duration for a qubit to retain classical and phase information, respectively. We model this error using the map $\mathcal{E}_{T_1,T_2}$, which is defined as the composition of the longitudinal $T_1$ relaxation followed by the transversal $T_2$ relaxation. For a duration $\Delta t$, the map is defined as
\begin{equation}
\label{eq:decay}
    \mathcal E_{T_1,T_2}(\Delta t) =  deph(\frac{1}{2}(1-e^{-\Delta t/T_2}))\circ gamp(1-e^{-\Delta t/T_1}), 
\end{equation}
where $gamp$ represents amplitude damping as defined in~\eqref{eq:gdamp}, and $deph(\delta)$ is a phase flip channel with the following map: 
\begin{equation}\label{eq:decay_sub_eq}
deph(\delta)(\rho)=(1-\delta)\rho + \delta Z\rho Z,
\end{equation}
where $\delta\in[0,1/2]$. 
For sufficiently long durations, as $\Delta t \to \infty$, the map $\mathcal{E}_{T_1,T_2}$ asymptotically drives the system towards its thermal state $\rho_T$, as defined in~\eqref{eq:rhot}. To account for this behavior, we apply the decay errors $\mathcal{E}_{T_1,T_2}$ to the qubits each time they remain idle. 

\paragraph{Measurement error}

Measurement errors are modeled classically, as the post-measurement quantum state is not utilized. Since qubits are typically reset after measurement, their final state does not influence subsequent operations.


The qubit readout error is captured with single-qubit confusion matrices, 
\begin{equation} \label{eq:readout}
    C_j = 
    \begin{bmatrix} 
    P_j(0\mid0) &  P_j(0\mid1) \\
    P_j(1\mid0) &  P_j(1\mid1) 
    \end{bmatrix},
\end{equation}
where $P_j(k\mid l)$ indicates probability of observing output $k$ on the prepared state $\ket l$ on qubit $j$. This model requires characterization of the matrix $C_j$ for each qubit.


\paragraph{Single-qubit gate noise}

The native single-qubit gates comprise the rotations $Rx(\theta), Ry(\theta)$, and $Rz(\theta)$, for an angle $\theta$. The $Rz(\theta)$ gates are implemented virtually by tracking the phase in the software, following the approach described in reference~\cite{helsen2017representation}. This technique is widely used in superconducting qubit platforms. Thus, all operators $Rz$ are assumed to be perfect and instantaneous.

We model the noise of $R_x$ and $R_y$ gates letting a dephasing error act after the original gate: 
\begin{equation} \label{eq:rot_error}
    deph(\delta_1).
\end{equation}  
The parameter $\delta_1$ is estimated from the average single-gate fidelity $\bar{F}$, given by  
\begin{equation} \label{eq:delta1}
    \delta_1 = \frac{3}{2} (1 - \bar{F}),~\text{and}~ \delta_1 \in \left[ 0, 0.5 \right].
\end{equation}
The value of $\bar{F}$ can be obtained from (interleaved) randomized benchmarking (RB)~\cite{Knill_2008, magesan2012efficient}. The derivation of Eq.~\eqref{eq:delta1} can be found in Appendix B of~\cite{Gustiani2025virtualquantum}. 

The exponential decay noise is parameterized by the gate duration. We assume a constant duration for all $\theta$ as rotation gates are implemented using calibrated microwave pulses with appropriate amplitude and fixed duration. In this model, the predominant noise is captured by dephasing, which accounts for random fluctuations in the qubit frequency commonly occurring during gate operations.


\paragraph{Two-qubit gate noise}
The native two-qubit gate in this device is the CZ gate. The gate is implemented by detuning the qubits to a common frequency, specifically by lowering the frequency of the higher-frequency qubit to match that of the lower-frequency qubit in the pair. The noise is modeled with 2-qubit dephasing noise $deph2$ with
the following map,

\begin{equation}\label{eq:dephase2}
deph2(\delta_2)(\rho)=(1-\delta_2)\rho + \frac{\delta_2}{3}(Z_j\rho Z_j+Z_k\rho Z_k+Z_jZ_k\rho Z_j Z_k).
\end{equation}

where $\delta_2\in[0,\frac{3}{4}]$, where in this case,

\begin{equation}\label{eq:delta2}
    \delta_2=5(1-\bar F_2)/4.
\end{equation}
Variable $\bar F_2$ indicates the average fidelity of the CZ gate, which can be obtained from interleaved randomized benchmarking~\cite{magesan2012efficient}. The derivation of Eq.~\eqref{eq:delta2} can be found in Appendix B of~\cite{Gustiani2025virtualquantum}.

We model the error as pure dephasing, which accounts for the reduction in the effective $T_2$ time due to the detuning of the higher-frequency qubit toward the lower one during gate execution. This choice is further supported by findings that incoherent errors are the dominant contribution to the overall error in CZ gate implementations~\cite{sete2024errorbudgetparametricresonance}.

\paragraph{Crosstalk noise}

We model crosstalk noise as an always-on ZZ interaction between neighboring qubits, as commonly observed in transmon devices~\cite{zhao2021suppression,Rudinger_2021,mckay2019three}. This interaction arises from cross-Kerr couplings, which result from higher-order perturbative effects in the dispersive regime of circuit quantum electrodynamics. The corresponding Hamiltonian takes the form
\begin{equation} \label{eq:always_on}
   H_{ZZ}=\beta Z_uZ_v,\quad\text{with}\quad\beta=J^2(\frac{1}{\Delta-\alpha_u}-\frac{1}{\Delta-\alpha_v}),
\end{equation}
where $\Delta = f_u - f_v$ is the qubit frequency detuning (with $f_u > f_v$), and $\alpha_u$, $\alpha_v$ are the anharmonicities of qubits $u$ and $v$, respectively. The quasi-coupling strength $J$ is fitted using experiments.


\begin{figure}[h]
    \centering
    \begin{tikzpicture}
        \node[circle,draw,fill=white!20,minimum size=1cm] (q2) at (0,0) {$Q_2$};
        \node[circle,draw,fill=white!20,minimum size=1cm] (q0) at (-2,2) {$Q_0$};
        \node[circle,draw,fill=white!20,minimum size=1cm] (q1) at (2,2) {$Q_1$};
        \node[circle,draw,fill=white!20,minimum size=1cm] (q3) at (-2,-2) {$Q_3$};
        \node[circle,draw,fill=white!20,minimum size=1cm] (q4) at (2,-2) {$Q_4$};
        
        \draw[-] (q2) -- (q1);
        \draw[-] (q2) -- (q0);
        \draw[-] (q2) -- (q3);
        \draw[-] (q2) -- (q4);
    \end{tikzpicture}
    \caption{Topology of the QPU with five qubits arranged in a star configuration.}
    \label{fig:qpu_topology}
\end{figure}
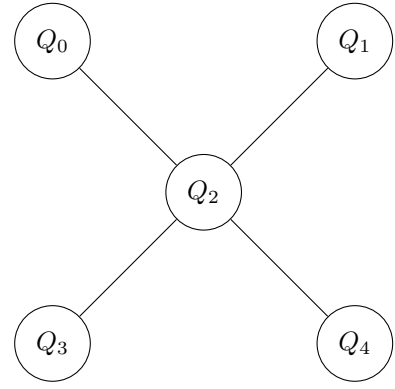

\section{Implementation of the noise model} \label{sec:noise_imple}

\noindent The noise model is integrated within a quantum emulation workbench~\cite{jennings2024quantum}, which serves as an emulation framework for quantum computing. This framework provides a digital twin of the quantum device, allowing for augmentation and continued development. It features a modular architecture that enables seamless integration with third-party software backends. In our approach, we employ a custom transpilation of the noise model within this emulation workbench, translating it into a representation compatible with the Qiskit Aer simulator~\cite{qiskit2024} for simulating noisy quantum circuits. The noisy circuit is subsequently evaluated using Qiskit's native density matrix simulation method.

We implement our noise model by incorporating noise instructions into the quantum circuits of interest. We distinguish two major categories of errors in the context of noise transpilation: \textit{instruction-based} and \textit{layer-based} errors. Instruction-based errors are applied at the gate level and can be conditioned on specific gate properties, such as fidelity. In contrast, noise that is applied at the layer level can be adapted to the layer duration, the coupling architecture, and/or to the occurrence of instructions inside the layer.

Instruction-based error rules are specified for each of the basis gates corresponding to the hardware model. Layer-based errors, on the other hand, are usually time-dependent passive errors that are applied per layer as a discrete approximation of their continuous effect throughout circuit execution.

For a given (transpiled) circuit intended for simulation within the noise model, a schedule incorporating all gate operations is first constructed, following an as-late-as-possible scheduling approach by default. While gate durations are assumed to be known, this scheduling process facilitates the extraction of any additional timing and duration information required for accurate noise modeling.

To apply the defined error rules, an intermediate directed acyclic graph (DAG) representation is utilized, allowing the insertion of noise operations at appropriate points. The quantum circuit is partitioned into sequential layers, with each layer containing at most one instruction per qubit. For a given qubit $j$ in layer $i$, we define its duration as $\Delta t_{ji}$, while the idle time of the qubit within the layer is denoted by $\Delta \epsilon_{ji}$. This per-qubit distinction is essential because, although the layer representation provides a structured framework with logical operators and discrete time steps, it does not inherently ensure that all qubits experience the same physical duration within a given logical layer.

We begin with an overview of the instruction-based errors associated with the basis gates used in our model, with the example set of basis gates given by \{Rx, Ry, Rz, CZ\}. These errors are tailored to match the quantum device described in Sec.~\ref{sec:qpu}, but the simulation framework allows for easy adaptation to different parameters and error rules as needed.

\begin{itemize}
    \item \textbf{Rx \& Ry}: The noise of these rotation gates is modeled by applying Eq.~\eqref{eq:rot_error}. The dephasing error is applied after the gate with parameter $\delta_1$, obtained from the results of RB, see Eq.~\eqref{eq:delta1}. 
    \item \textbf{Rz}: The Rz rotation is a virtual rotation in the analyzed architecture and is assumed to be error free. Therefore, no error is added for Rz gates and its duration is 0 ns.
    \item \textbf{CZ}: The CZ gate error is applied after the entangling gate. First, the dephasing noise  is applied according to the parameter $\delta_2$ which can be obtained from RB or fitted via optimization, see Eq.~\eqref{eq:dephase2} and~\eqref{eq:delta2}.    
    \item \textbf{Readout error}: The readout error is applied after each measurement gate as a fully classical operation. It applies the bit-flip maps $\{ C_j \}$ for each qubit, see Eq.~\eqref{eq:readout}. The transition probabilities are assumed to be known from respective characterization experiments during the calibration process.
\end{itemize}

 \noindent Additionally, our model deploys three layer-based noise channels: 

\begin{itemize}
    \item \textbf{State preparation error}: The state preparation error is applied prior to the first layer of quantum instructions. It is modeled using the generalized amplitude damping channel, as described in Eq.~\eqref{eq:gdamp}. The excited state probability $p_j^1$ is extracted from the calibration data and is specific to each qubit. For our simulations, we set the dampening strength $\gamma = 1$, and an active reset initialization was used~\cite{PhysRevLett.109.240502, Magnard_2018}.
    
    \item \textbf{Decay}: For each layer $i$, a decay noise is applied on  qubit $j$ via the map $\mathcal{E}_{T_1,T_2}$ for the duration the qubit is idle, $\Delta \epsilon_{ji}$, see Eq.~\eqref{eq:decay}. In the simulations, $\gamma=1-e^{-\Delta t/T_1}$, the excited state probability $p_j$ is taken from the calibration, and $\delta$ is set to 0.5, see Eq.~\eqref{eq:decay_sub_eq}. If required, the decay noise is applied inside a layer, both before and after the original instructions. Scenarios that require this handling appear as layers can have a carry-over from the previous layer, and gates inside a single layer can have different durations. Notably, we consider the passive decay during the gate execution to be captured by the error of the gate itself.
    
    \item \textbf{Crosstalk}: In our model, an always-on coupling error~\cite{zhao2021suppression,Rudinger_2021,mckay2019three} is used as the main crosstalk error. The crosstalk error is applied as a two-qubit operation at the end of each layer, given by 
    \begin{equation} \label{eq:crosstalk_explicit}
        CT(d_{uv}) = \exp(-i\beta d_{uv}), 
    \end{equation}
    
    with $\beta$ defined in Eq~\eqref{eq:always_on}. This operator acts diagonally in the computational basis. It is applied pair-wise between all direct neighbors of qubits that take part in the quantum circuit, i.e., including inactive neighboring qubits that are on the device but not actively used in the current quantum circuit. The duration of the error is chosen as the maximum layer duration of the participating qubits, i.e., $d_{uv, i}=\max(\Delta t_{ui}, \Delta t_{vi})$ for participating qubits $u$ and $v$ in layer $i$. For our simulations, we parametrize this error by treating the coupling constant $J$ as a variable value that can be optimized during the fitting of the model. If a qubit is coupled to multiple qubits, the application order is determined heuristically based on the numerical labels of the respective qubits.
\end{itemize}

\begin{figure}
    \centering
    \begin{quantikz}
    &\gate{R_x(\theta)} &   \hspace{0.3cm} \Rightarrow  \hspace{0.3cm} & \gate{R_x(\theta)} & \gate{deph(\delta_1)} \\
    &\gate{R_y(\theta)} &   \hspace{0.3cm} \Rightarrow  \hspace{0.3cm} & \gate{R_y(\theta)} & \gate{deph(\delta_1)} \\
    &\gate{R_z(\theta)} &   \hspace{0.3cm} \Rightarrow  \hspace{0.3cm} & \gate{R_z(\theta)} &  \qw  \\
    \end{quantikz}
    \caption{Noise rules for single-qubit gates. Definitions of the noise instructions can be found in Sec.~\ref{sec:noise_model_theo}.}
    \label{fig:active_noise_single}
\end{figure}
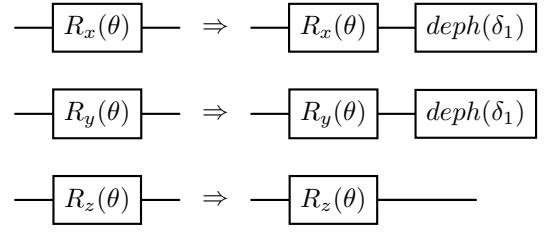

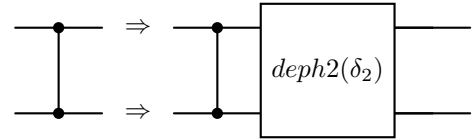
\begin{figure}
    \centering
    \begin{quantikz}
    &\ctrl{1} &    \hspace{0.3cm}\Rightarrow  \hspace{0.3cm}  & \ctrl{1} &  \gate[2]{deph2(\delta_2)} & \qw &\\
    &\control{} &   \hspace{0.3cm}\Rightarrow  \hspace{0.3cm} & \control{} & & \qw &
    \end{quantikz}
    \caption{Noise rule for the CZ gate, the entangling gate used in the gate set. Definitions of the noise instructions can be found in Sec.~\ref{sec:noise_model_theo}.} 
    \label{fig:active_noise_twice}
\end{figure}

\begin{figure}
    \centering
    \begin{quantikz}
    \lstick{$q_0$} &\gate[3]{CT(d_{02})}\gateinput{0}   &&&\\
    \lstick{$q_1$} & &                                       \gate[2]{CT(d_{12})}\gateinput{0} && \\
    \lstick{$q_2$} & \gateinput{1}                           &\gateinput{1}  & \gate[2]{CT(d_{23})}\gateinput{0} & \gate[3]{CT(d_{24})}\gateinput{0} \\
    \lstick{$q_3$} & &&\gateinput{1}  &\\
    \lstick{$q_4$} &&&&\gateinput{1}  
    \end{quantikz}
    \caption{The crosstalk instructions that are applied after each layer. The QPU has a star architecture, with qubit $q_2$ in the center, see Fig.~\ref{fig:qpu_topology}. The order of application is heuristic, the instruction itself is given by Eq.~\eqref{eq:crosstalk_explicit}, and the calculation of the applicable duration $d_{uv}$ is described in Sec.~\ref{sec:noise_imple}.}
    \label{fig:passive_CT}
\end{figure}
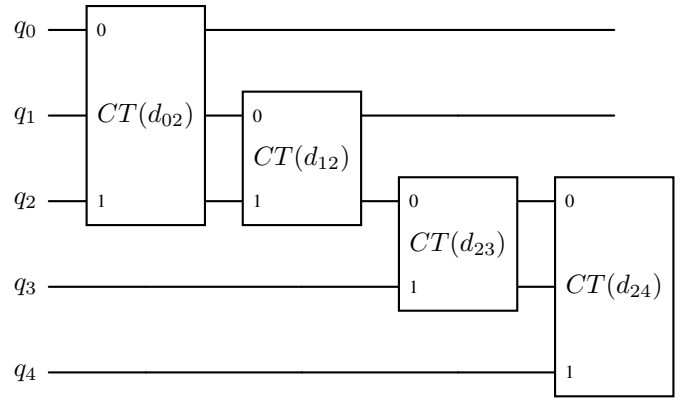


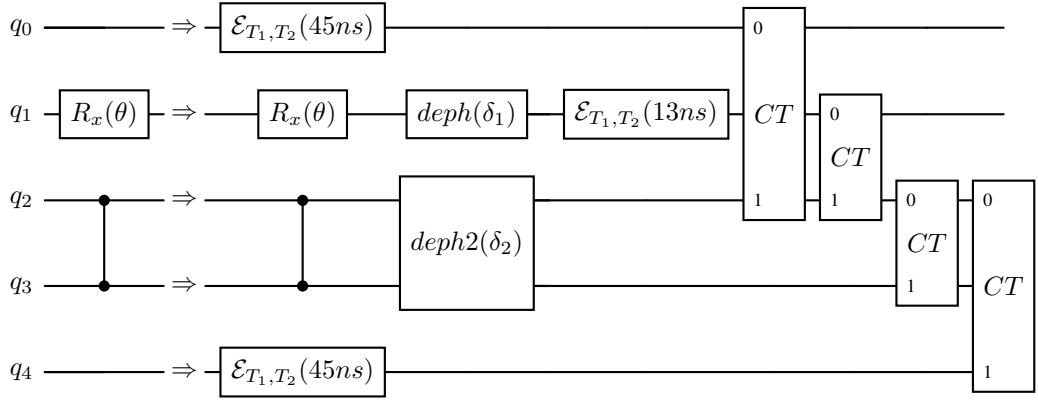
\begin{figure*}
    \centering
    \begin{minipage}{\textwidth}
        \centering
        \begin{quantikz}[column sep=0.2cm]
        \lstick{$q_0$} & \qw &  \hspace{0.1cm}\Rightarrow\hspace{0.1cm}
        &\gate{\mathcal{E}_{T_1,T_2}(45ns)} & & &
        &\gate[3]{CT}\gateinput{0}   &&&\\
        \lstick{$q_1$} & \gate{R_x(\theta)} & \hspace{0.1cm}\Rightarrow\hspace{0.1cm}
        &\gate{R_x(\theta)} &\gate{deph(\delta_1)} & \qw & \gate{\mathcal{E}_{T_1,T_2}(13ns)}
        & &                                       \gate[2]{CT}\gateinput{0} && \\
        \lstick{$q_2$} & \ctrl{1} & \hspace{0.1cm}\Rightarrow\hspace{0.1cm}
        & \ctrl{1} & \gate[2]{deph2(\delta_2)} & \qw &
        & \gateinput{1}                           &\gateinput{1}  & \gate[2]{CT}\gateinput{0} & \gate[3]{CT}\gateinput{0} \\
        \lstick{$q_3$} & \control{} &  \hspace{0.1cm}\Rightarrow\hspace{0.1cm}
        & \control{} & & \qw &
        & &&\gateinput{1}  &\\
        \lstick{$q_4$} & \qw & \hspace{0.1cm}\Rightarrow\hspace{0.1cm}
        & \gate{\mathcal{E}_{T_1,T_2}(45ns)} & & &
        &&&&\gateinput{1}  
        \end{quantikz}
    \end{minipage}
    \caption{Example of noise transpilation of a single layer. We assume a CZ gate duration of 45 ns and a X-rotation gate time of 32 ns. All crosstalk in this layer is applied for 45 ns.}
    \label{fig:transpilation}
\end{figure*}

\begin{figure*}
    \centering
    \includegraphics[width=0.75\linewidth]{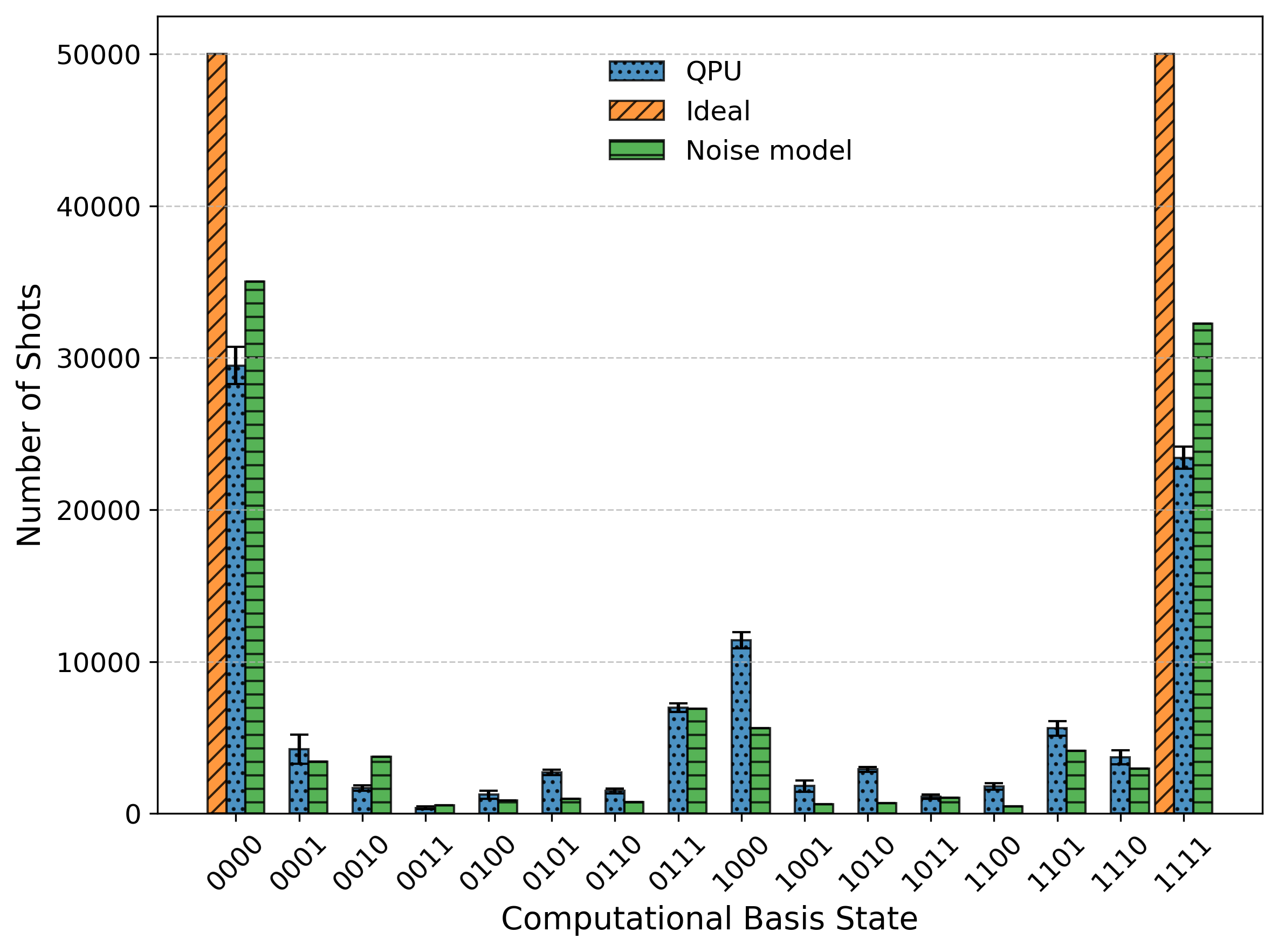}
    \caption{The measurement outcomes of a 4-qubit GHZ circuit measured in the computational basis. Ideal refers to the expected outcome of the state, $1/\sqrt{2}(\ket{0000}+\ket{1111}$. QPU refers to the outcomes of circuits run on the device, with the bar height being the mean of 80 circuits run over a period of 33 hours at 100.000 shots each. The error bars correspond to one standard deviation. The noise model bar shows the result of the simulated noise model, the error bars correspond to the statistical error from using different seed values. The mean TVD value of this test circuit is $0.18$.}
    \label{fig:shots}
\end{figure*}

\begin{figure*}
        \centering
        \includegraphics[width=0.75\linewidth]{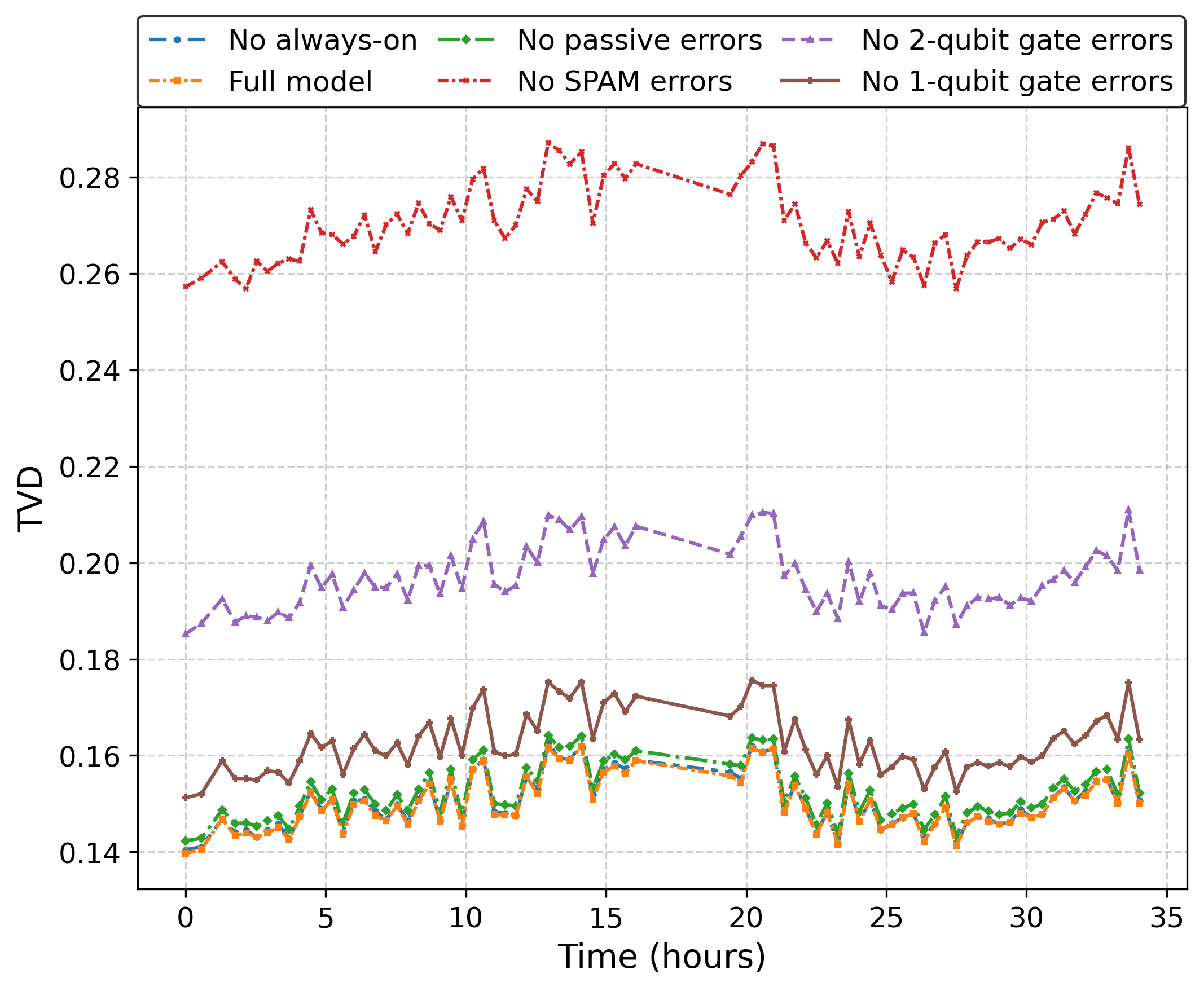}
        \caption{Mean TVD between QPU and noise model shot distributions of 8 benchmarking circuits over time since the last calibration. The different lines correspond to the full noise model and to specific effects being turned off in emulation. A detailed description of which labels correspond to which errors being toggled off can be found in Sec.~\ref{sec:exp_res}. Note that the "No Always-on" line is partially covered by the curve of the full model. The empty patch around hour 17 was the result of a temporary lost connection to the QPU. The mean TVD value over all data of the full model is 0.15.}
        \label{fig:tvd}
\end{figure*}

\noindent A visualization and examples of the noise transpilation can be seen in Fig.~\ref{fig:active_noise_single},~\ref{fig:active_noise_twice},~\ref{fig:passive_CT}, and~\ref{fig:transpilation}. The order in which we apply the different errors is arbitrary; however, it can generally have a substantial effect on quantum simulations, as for example demonstrated in~\cite{Tranter_2019}. To the best of our knowledge, an analysis and optimization of the order of errors in this type of noise model remains an open question for further research. For the simulations done in this paper, the single coupling constant $J$ and the $n-1$ fidelities of the CZ gates are chosen as parameters that are free to be optimized during fitting, resulting in a total of $n$ free parameters, where $n$ denotes the number of qubits.


\section{The supercoducting qubit device} \label{sec:qpu}

\noindent We evaluate our noise model on a five-qubit quantum processing unit (QPU), specifically the \textit{Soprano-D} QPU from QuantWare, featuring flux-tunable transmon qubits. The qubits are arranged in a star topology with fixed couplers, as shown in Fig.~\ref{fig:qpu_topology}. The native gate set consists of $\{\text{Rx}, \text{Ry}, \text{Rz}, \text{CZ}\}$, where the $\text{Rz}$ gate is implemented as a virtual gate. Due to high readout error on qubit $Q_4$, we restrict our experiments to the first four qubits. However, since $Q_4$ remains coupled to the active qubits, we include it in our model to account for its always-on crosstalk, particularly affecting qubit $Q_2$. The measured qubit frequencies range from $4.5$ GHz to $6.3$ GHz, increasing with the qubit index. The relaxation ($T_1$) and dephasing ($T_2$) times are measured to be between 32–45 $\mu$s and 14–21 $\mu$s, respectively. The qubit anharmonicities are approximately $200$ MHz. The average readout error across the active qubits is $3.3\%$, while the average state preparation error is $4.8\%$. Gate fidelities are quantified through randomized benchmarking~\cite{magesan2012efficient}, demonstrating an average single-qubit gate fidelity of $0.996$ for all non-virtual basis gates, each executed within $32$ ns. The two-qubit CZ gate employs dynamic frequency tuning via flux-bias control of the higher-frequency qubit, aligning it to the resonance frequency of the target qubit, achieving a gate duration of $45$ ns. Measurement pulses have a duration of $1500$ ns.

\section{Experimental results} \label{sec:exp_res}

\noindent We train and test the performance of the noise model using fundamental physics benchmarking circuits, i.e., GHZ- and W-state preparation circuits~\cite{greenberger2007goingbellstheorem, D_r_2000}, following the example of~\cite{PhysRevA.103.042603, abdurakhimov2024technologyperformancebenchmarksiqms}. The used circuits are run on the quantum device described in Sec.~\ref{sec:qpu}, where we vary the circuits over different combinations on 3 or all of the 4 active qubits. To ensure the resilience and longevity of the digital twin, we collect data in clusters of 8 benchmarking circuits over a duration of 33 hours, with breaks between the clusters of 15 to 30 minutes. Each circuit is evaluated with 100.000 shots to minimize the statistical noise. The full model has $n=4$ free parameter, i.e., the coupling strength $J$ of the static crosstalk noise and the fidelities of the CZ gates. We use only a small fraction of the collected data ($12.5\%$) and only 3-qubit circuits to optimize the model while the rest and the 4-qubit circuits are used to test the performance of the predictions. Differential Evolution~\cite{Storn1997} is used to optimize the model's parameters, with the cost function being the total variational distance (TVD)~\cite{rachev1998mass} of the shot distributions. From the best-performing candidates, we holistically selected one that fits the physical expectations for the device for the rest of the analysis. The respective parameters can be seen in Tab.~\ref{tab:params} and result in a mean TVD of 0.15 on all test circuits. The fidelity and cross-talk interaction strength differ from the values expected from a purely physical analysis. This freedom was intentionally granted to the optimizer, allowing these parameters to account for errors that are not captured by our model. An example of the predicted shot distribution can be seen in Fig.~\ref{fig:shots}, where a 4-qubit GHZ state is prepared between qubits 0, 1, 2, and 3. The QPU bar shows the mean results of all the respective state-preparation circuits, while the noise model bar shows the prediction of the emulation, all for measurements in the Z-basis. The TVD on this specific test circuit shows only a slight degradation to a TVD of 0.18 despite the increase in size. Fig.~\ref{fig:tvd} illustrates the TVD distance between the emulation and the actual QPU over the time passed since the last calibration, including all captured data. The different curves correspond to different error channels being turned off to visualize their impact. There, "No passive errors" corresponds to the exponential decay error, Eq.~\eqref{eq:decay}, being toggled off. "No SPAM errors", i.e., no state preparation and measurement (SPAM) errors, relates to the reset and measurement error being turned off, Eq.~\eqref{eq:rhot} and Eq.~\eqref{eq:readout}, respectively. "No 2-qubit gate errors" and "No 1-qubit gate errors" correspond to Eq.~\eqref{eq:rot_error} and Eq.~\eqref{eq:dephase2}, respectively. Finally, "No Always-on" represents a toggle-off of the crosstalk interaction, Eq.~\eqref{eq:crosstalk_explicit}.



\begin{table}[]
    \centering
    \caption{Optimized noise model parameters for the tested QPU. The fidelity corresponds to $1-\delta_2$ in Eq.~\eqref{eq:dephase2} and the Always-on interaction strength corresponds to $J$ in Eq.~\eqref{eq:always_on}.}
    \begin{tabular}{|c|c|c|c|}  
        \hline
        \multicolumn{4}{|c|}{CZ Fidelity ($q_i-q_j$)}\\
        \hline
         0-2 & 1-2  & 3-2 &  4-2\\  
         \hline
         0.987& 0.964 & 0.917& --\\
         \hline
         \multicolumn{4}{|c|}{Always-on interaction-strength [kHz]}\\
         \hline
         15.0 & 22.2 & 19.2 & 65.7\\
         \hline         
    \end{tabular}

    \label{tab:params}
\end{table}

\section{Discussion}


\noindent The mean TVD on all collected data evaluates to a mean TVD of 0.15, the TVD between the QPU results and an ideal outcome is about twice that value. This is slightly worse but in a similar range as comparable approaches in~\cite{PhysRevA.103.042603, 9193969, weber2023volumetricbenchmarkingquantumcomputing}, which all use fixed-frequency transmon devices by IBM compared to the tunable-frequency transmon device analyzed here. The model in~\cite{PhysRevA.103.042603} is characterized using an array of GHZ-states, the model is not a physics-inspired model but is instead based only on the characterization of gate errors, and we therefore expect it to require more data to fit compared to our physics-inspired noise model. The model in~\cite{9193969} additionally uses the same parameterized crosstalk as we use, but relies on running Idle Tomography~\cite{Blume-Kohout2019} and Simultaneous Randomized Benchmarking~\cite{Gambetta_2012} to fit their parameters. Reference~\cite{weber2023volumetricbenchmarkingquantumcomputing} employs a noise model that overlaps with our proposed model in all but the 2-qubit noise, where they use conditional rotations instead. As a clear distinction, they do not use any physical properties but fully learn a parametrization of all errors, resulting in roughly 11 times the amount of parameters compared to our model. In contrast, our hybrid approach uses a strongly physics-inspired noise model that only requires basic physical calibration information and randomized benchmarking in combination with $n$ parameters to be fitted by using simple application-inspired circuits. Notably, our noise model stands as one of the few that are tested on a superconducting quantum device that is not fabricated by IBM.\\

\noindent While a common approach is to extract the 2-qubit gate error rate via randomized benchmarking or its variants, we instead chose to extract those rates through fitting of the noise model on simple but expressive circuits. This is based on the heuristic approach that this freedom allows the optimizer to catch any error sources that are not explicitly implemented in our noise model. Following a similar train of thought, we parametrized the always-on interaction strength via a variable coupling strength $J$. The resulting interaction strengths, see Tab.~\ref{tab:params}, are slightly below the interaction strengths one would obtain when using the fixed coupling strength provided by the QPU fabricator. The biggest contributors to the noise are SPAM and gate errors while activating the passive and crosstalk error channels only slightly increases the performance of the noise model. We expect them to significantly increase in impact on setups that have lower gate and SPAM error rates, allowing for the execution of longer circuits. Nevertheless, the data suggest that other noise channels not included in our model contribute significantly. Possible candidates are pulse spillover,  classical line crosstalk, or flux fluctuations~\cite{Dai_2021, heng2024estimatingeffectcrosstalkerror, Sarovar_2020}. Additionally, the remaining TVD distance could also hint at the presence of strong non-Markovian errors~\cite{Rudinger_2021}.

\section{Conclusion}

\noindent We present a parametric error model integrated into a modular quantum emulation framework. The noise model of the digital twin is strongly physics-inspired, specifically tailored for a flux-tunable transmon quantum processing unit. By leveraging calibration data acquired during each calibration run, we create and maintain a digital representation that adapts to variations and drifts in the device over time, successfully accounting for these changes.

Our noise model offers high expressibility while requiring only $n$ parameters, which can be efficiently fitted using specific benchmarking circuits. In our case, a simple combination of GHZ- and W-state preparation proves to be effective. We demonstrate the capabilities of the digital twin on a 5-qubit device, showing that it can predict shot distributions with high accuracy, achieving a mean total variation distance (TVD) of 0.18 on an extended test circuit. Furthermore, for the tested QPU, the digital twin reliably predicts quantum circuit outcomes after more than 30 hours. We analyze the contributions of different error channels and demonstrate a versatile approach to digital twins for transmon qubits. Our method provides strong emulation capabilities, maintains precision through calibration-dependent adjustments, enables resource-efficient parameter fitting, and delivers highly interpretable results.






\section*{Acknowledgment}
\noindent We gratefully acknowledge the support provided by the CQC facility at the Jülich Supercomputing Center. We particularly thank David Rabanus and Eduardo Schätzle for their outstanding engineering work and dedicated efforts in setting up and maintaining the experimental infrastructure used in this study.

\bibliographystyle{IEEEtran}
\bibliography{main.bib}
\end{document}